# Electrical injection and detection of spin accumulation in Ge at room temperature

A.T. Hanbicki[a,*], S.-.F. Cheng[a], R. Goswami[b], O.M.J. van 't Erve[a], and B.T. Jonker[a]

[a]*Naval Research Laboratory*, 4555 Overlook Ave. SW, Washington, DC 20375

[b]*SAIC Inc.*, 1710 SAIC Dr., Mclean, VA 22012

ABSTRACT

We inject spin-polarized electrons from an Fe/MgO tunnel barrier contact into *n*-type Ge(001) substrates with electron densities $2 \times 10^{16} < n < 8 \times 10^{17}$ cm$^{-3}$, and electrically detect the resulting spin accumulation using three-terminal Hanle measurements. We observe significant spin accumulation in the Ge up to room temperature. We observe precessional dephasing of the spin accumulation (the Hanle effect) in an applied magnetic field for both forward and reverse bias (spin extraction and injection), and determine spin lifetimes and corresponding diffusion lengths for temperatures of 225 K to 300 K. The room temperature spin lifetime increases from $\tau_s = 50$ ps to 123 ps with decreasing electron concentration, as expected from electron spin resonance work on bulk Ge. The measured spin resistance–area product is in good agreement with values predicted by theory for samples with carrier densities below the metal-insulator transition (MIT), but $10^2$ larger for samples above the MIT. These data demonstrate that the spin accumulation measured occurs in the Ge, although dopant-derived interface or band states may enhance the measured spin voltage above the MIT. We estimate the polarization in the Ge to be on the order of 1%.

* Corresponding Author. Tel.: (202) 767-3956; fax: (202) 404-4637.
*E-mail address*: Aubrey.Hanbicki@nrl.navy.mil

# 1. Introduction

Electrical injection and detection of spin accumulation in Si at temperatures up to 500 K [1] has demonstrated that spin angular momentum is a promising alternate state variable for practical applications, and provides an avenue for introducing new functionality into electronic devices [2,3,4]. While significant progress has been made in Si since 2007, [5,6,7,8,9] research on spin transport in Ge is still in a nascent stage. Ge is attractive as a *p*-channel replacement material for standard complementary metal-oxide-semiconductor (CMOS) applications due to its high hole mobility [10]. As a potential host for spin transport, its strong spin-orbit interaction may enable electric field control of spin precession needed for realization of a spin field effect transistor [4,11] in a group IV-based device.

Several recent reports have described spin effects in Ge [12,13,14,15,16,17]. Spin-polarized electrons were optically generated in Ni/Ge/AlGaAs/GaAs heterostructures in one or more of the semiconductor layers, and detected as a photocurrent at the surface Ni contact which served as a spin analyzer [12]. Although the high incident photon energy (1.96 eV, He-Ne laser) will excite only very small spin polarization in AlGaAs/GaAs [18], and even less in Ge [19], helicity dependent effects were observed to 300 K.

Other authors [13-16] used the Hanle effect in a three-terminal geometry (described below) to electrically detect a voltage attributed to spin accumulation either in the Ge or on interface states. Spin accumulation was reported in *p*-Ge up to 100 K with a hole spin lifetime of 30 ps, although an unexplained $10^3$ enhancement of the spin signal over the theoretical value was observed [13]. Jain, *et al.* also observed an unexpected $10^4$ enhancement of the spin voltage in *n*-Ge up to 220 K with an electron spin lifetime of 35 ps [14], which they attributed to spin accumulation on localized interface states as proposed by Tran, *et al.* [20]. Other authors report room temperature spin lifetimes in *n*-Ge ranging from 100~120 ps [15,16]. In all of these three-terminal Hanle studies, spin accumulation was measured for a single Ge substrate doping, and no attempts were made to correlate the observed spin lifetimes with a change in the properties of the semiconductor substrate to definitively show that the measured voltage originated from spin accumulation in the Ge rather than in interface states. While the spin voltage data of the three-terminal geometry may provide valuable insight into spin accumulation in the semiconductor channel, such data must incorporate variables that show clear correlation with the semiconductor.

Zhao, *et al.* used the four-terminal non-local spin valve geometry to unambiguously show the generation and detection of pure spin currents and corresponding spin transport in *n*-Ge up to 225 K [17].

We report here electrical injection and detection of spin accumulation in *n*-Ge(001) at room temperature using Fe/MgO/tunnel barrier contacts on a series of *n*-doped Ge substrates ($2\times10^{16} < n < 8\times10^{17}$ cm$^{-3}$). We observe precessional dephasing of the spin accumulation (the Hanle effect) in an applied magnetic field for both forward and reverse bias (spin extraction and injection), and determine spin lifetimes and corresponding diffusion lengths for temperatures of 225 K to 300 K. We find that the spin lifetime increases with decreasing electron concentration, as expected from electron spin resonance work on bulk Ge, and that the measured spin resistance–area product is in good agreement with values predicted by theory. These data demonstrate that the spin accumulation measured occurs in the Ge rather than in localized interface states.

## 2. Experimental

Samples were prepared from As-doped, epi-ready, *n*-type Ge(001) wafers. Three different substrates were used, and the room temperature carrier concentration (*n*), mobility, and resistivity ($\rho$) obtained with van der Pauw measurements are summarized in Table I. Substrates were inserted into a sputter deposition chamber and annealed at 580°C to remove the native oxide. After cooling, 10 Å films of MgO were sputter deposited at 140W from an MgO target in 3mTorr of Ar at substrate temperatures less than 40°C. A 120 Å layer of Fe was then sputter deposited at 120W in 2mTorr Ar at room temperature.

Figure 1(a) shows a transmission electron microscope (TEM) image of sample *A*. The MgO layer is amorphous and is 7-10 Å thick, while the 120 Å thick Fe film is composed of large polycrystalline grains. The magnetic behavior of the as-grown Fe films was measured in-plane at room temperature using the magneto-optic Kerr effect, and out-of-plane using a vibrating sample magnetometer. For all of the samples, the magnetization of the Fe is in the plane of the film and has a four-fold anisotropy, characteristic of bulk-single crystal Fe indicating that the Fe is a textured polycrystal, consistent with the TEM observations.

The samples were processed into devices suitable for transport measurements using standard photolithography and chemical etching techniques. Figure 1(b) shows a schematic diagram of the heterostructure and device geometry.

## 3. Results and discussion

Previous work has shown that MgO contacts can change the pinning of the Fermi energy [21,22,23] as well as provide effective tunnel transport, thus circumventing the conductivity mismatch which prevents efficient electrical spin injection from a ferromagnetic metal into a semiconductor [24,25]. Electrical characterization (*I-V*) using a 2-wire geometry was performed with the same 100 μm x 150 μm devices used for the spin accumulation measurements described below. *I-V* data are shown as a function of temperature (200-295 K) for sample *A* in Fig. 2(a) and at room temperature for samples *B* and *C* in Fig. 2(b). The samples exhibit a rectifying behavior typical of a Schottky contact rather than the symmetric behavior expected for an ideal tunnel junction. The predominantly Schottky character of the contact is attributed to the wide depletion layer in the Ge which exists at these carrier concentrations [26].

The *I-V* data were fit using the approach described in Ref. 27, where a standard diode equation is augmented with a series resistance and perturbative parallel transport path to model real systems. Fitting parameters include the Schottky barrier height, $\Phi_{SB}$, the ideality factor **n**, a series resistance $R_s$, and a parallel path conductance $G_p$. Excellent fits to the data are obtained, shown as solid lines on the *I-V* curves of Fig. 2, and Table I summarizes the parameters used. In our samples, the height of the Schottky barrier has been reduced from the fully pinned value of nearly 0.6 eV to an intermediate value near 0.35 eV because of the insertion of the thin MgO layer [21]. This lower barrier facilitates tunneling, and indeed the ideality factor of **n**=1.15 obtained for all the samples indicates that there are components to the current other than thermionic emission.

Spin accumulation and precession directly under the magnetic contact can be observed using the geometry shown in Fig. 1(b), which consists of three terminals: two independent reference contacts, and a central magnetic tunnel contact used as both spin injector and detector [1,8,20,28]. Individual devices were either reverse biased to inject electrons from Fe into Ge (spin injection), or forward biased to extract electrons from Ge into the Fe (spin extraction). The spin-polarized carriers produce a net spin accumulation in the Ge which results in a splitting of the electrochemical potential, $\Delta\mu=\mu_{up}-\mu_{down}$. This difference in chemical potential is related to a change in the voltage measured at the contact $\Delta V= \gamma \cdot \Delta\mu/2$ [25], where $\gamma$ is the tunneling spin polarization. The spin accumulation is reduced by an applied magnetic field, $B_z$, perpendicular to the electron spin direction, which produces precessional dephasing. The spins precess around $B_z$ at the Larmor frequency, $\omega_L = \frac{g\mu_B}{\hbar}|\vec{B}|$, and the net spin accumulation decreases to zero as $B_z$ increases (Hanle effect) [18]. Here, $g$ is the Landé $g$-factor ($g = 1.6$ for Ge [29]), $\mu_B$ is the Bohr magneton, and $\hbar$ is the reduced Planck's constant. The functional dependence of $\Delta V$ on the applied field is given by $\Delta V(B_z) = \Delta V_0 / (1+(\omega_L \tau_s)^2)$, a simple Lorentzian. Fitting parameters in this analysis include the magnitude of $\Delta V_o$ (and hence $\Delta\mu$), and the spin lifetime $\tau_s$.

Figure 3 summarizes Hanle effect data typical of data obtained from multiple devices. Fig. 3(a) shows data from sample $A$ at room temperature for currents corresponding to both electron spin injection (left axis) and extraction (right axis) for several bias values. In the extraction case, spin accumulation occurs because as majority electrons preferentially tunnel into the magnetic Fe contact, a minority spin polarization builds up in the Ge and the Hanle curve inverts, as expected. Note that there is a large difference between the magnitude of the Hanle signal in injection and extraction: $\Delta V_0$ is much smaller for biases corresponding to extraction. Data from sample $A$ for temperatures of 225-295 K at a fixed bias voltage of –0.5 V are presented in Fig. 3(b). Below 225 K the sample is too resistive to reliably measure. Room temperature Hanle data for samples $B$ and $C$ are shown in Figure 3(c) and (d), respectively, for several bias conditions corresponding to spin injection from the Fe into the Ge. Data points are open circles, and Lorentzian fits to the data as described above are shown as solid lines. A smoothly varying background has been removed from the data. As $B_z$ increases, precessional dephasing reduces the spin accumulation and the corresponding signal $\Delta V$ to zero, producing a characteristic lineshape which is very clear in the data for each of the three sample doping concentrations studied.

Control samples with non-magnetic tunnel barrier contacts were also fabricated and measured and exhibit no Hanle signal, ruling out anomalous effects such as magnetoresistance from the Ge. Measurements made with the magnetic field applied in-the-plane of the samples are dominated by background magneto-transport effects.

The magnitude of ΔV provides a measure of the polarization of the accumulated spin. At the largest injection currents used for each sample (shown in Figure 3 at 295 K, and limited by device failure), we measure $\Delta V_o \approx$ -150 µV, -8 µV, -40 µV for samples *A*, *B*, and *C*, respectively. Using $\gamma$ =0.4 [30], these correspond to $\Delta\mu$ = 0.75 meV, 0.05 meV, and 0.20 meV. Assuming a parabolic conduction band and Fermi-Dirac statistics, it is possible to then calculate the density of spin-up and spin-down electrons, $n_{up}$ and $n_{down}$, to determine the polarization of the accumulated spins, P = ($n_{up}$ – $n_{down}$ )/ ($n_{up}$ + $n_{down}$). Room temperature electron spin polarizations are 1.4%, 0.1% and 0.4%. This polarization is somewhat smaller than the polarization seen in room temperature spin accumulation experiments in Si [1,5,8], consistent with the larger spin-orbit interaction in Ge.

The spin lifetimes, $\tau_s$, obtained from the fits to the data are shown in Fig. 4 for T = 295 K as a function of electron density, and increase from 51±3 ps (sample *A*) to 88±14 ps (sample *B*) to 123±10 ps (sample *C*) as the electron density decreases. This trend is consistent with previous electron spin resonance (ESR) measurements on bulk Ge [31]. The lifetimes are slightly longer for spin extraction – for sample *A* we measure $\tau_s$ = 67±10 ps. For comparison, spin lifetimes obtained from ESR measurements of electrons in bulk Ge at similar electron densities and 80 K are ~2 ns [31]. This difference can be due to a variety of effects and conditions including temperature and the local environment where the spin lifetime is measured. The ESR work has shown that the spin lifetime increases significantly with decreasing temperature below 80 K for the range of doping concentrations we study, although the behavior at higher temperatures has not been addressed [31]. In our surface contact geometry, the spins are detected directly below the magnetic contact, and are therefore subject to reduced symmetry and increased scattering from the surface or near surface impurities, which will reduce the lifetime [1,8]. As the magnetic contact interface is likely to introduce additional scattering and spin relaxation mechanisms not present in the bulk, the region of the semiconductor directly beneath the contact is expected to be a critical factor in any future spin technology.

The spin diffusion length can be obtained from the spin lifetime as $\lambda_s = \sqrt{D\tau_s}$, where *D* is the diffusion constant calculated using Fermi-Dirac integrals and the measured carrier concentration and mobility [26]. For these devices, the average $\lambda_s$ = 0.37 µm, 0.52 µm, and 0.83 µm for samples *A–C* (Fig. 4, triangles, right axis). Therefore, while the spin polarization and lifetime in these Ge devices are smaller than the corresponding values measured in Si at room temperature [1], the larger mobility in Ge leads to a significantly longer spin diffusion length.

The spin signal predicted by the theory of diffusive transport across a single ferromagnetic metal/semiconductor interface for the geometry employed here is given by $\gamma^2 r_1 = \gamma^2 (\rho \cdot \lambda_s)$ [32,33]. For our samples $A$–$C$, the corresponding values are 5 Ω-μm$^2$, 15 Ω-μm$^2$, and 212 Ω-μm$^2$, respectively. The spin signals we measure experimentally for the largest injection currents used for each sample, given by the spin-resistance area product $\Delta V_o \cdot A/I$, where $A$ is the contact area and $I$ is the bias current, are 750 Ω-μm$^2$, 6 Ω-μm$^2$, and 60 Ω-μm$^2$ for samples $A$-$C$ at room temperature. The experimental values for samples $B$ and $C$ are in excellent agreement with theory, while the value for sample $A$ is about 150 times higher than the theoretical value. The electron concentration of sample $A$ is well above the metal-insulator transition (~$3 \times 10^{17}$cm$^{-3}$) [34], so that spin accumulation in dopant-derived band states may play a role in enhancing the spin voltage. Further work is underway to account for this discrepancy.

## 5. Summary

In summary, we have demonstrated spin accumulation in Ge(001) at room temperature by showing (a) a correlation between the measured spin lifetime and the Ge electron density, consistent with previous ESR measurements, and (b) that the magnitude of the measured spin voltage is in good agreement with theory for samples with carrier concentrations below the metal-insulator transition. We estimate the spin polarization in the Ge to be on the order of 1%.

**Acknowledgments**
We would like to acknowledge C.H. Li and G. Kioseoglou for useful conversations. This work was supported by core programs at NRL.

TABLE I:

| Sample | doping – n cm$^{-3}$ | mobility cm$^2$/Vs | resistivity – $\rho$ $\Omega$-cm | $\Phi_{SB}$ eV | n | $R_s$ $\Omega$ | $G_p$ S |
|---|---|---|---|---|---|---|---|
| A | 8.4x10$^{17}$ | 875 | 8.6x10$^{-3}$ | 0.38 | 1.15 | 6.36 | 2x10$^{-3}$ |
| B | 3.4x10$^{17}$ | 1050 | 1.7x10$^{-2}$ | 0.32 | 1.15 | 6.1 | 4x10$^{-2}$ |
| C | 2.1x10$^{16}$ | 1875 | 0.160 | 0.35 | 1.15 | 11.1 | 8x10$^{-3}$ |

FIGURE CAPTIONS:

FIG. 1. (a) TEM image of the Fe/MgO/Ge heterostructure obtained from sample A. (b) Schematic diagram of the heterostructure and a typical 3-terminal device geometry.

FIG. 2. I-V characteristics of (a) sample A from 200 K to 295 K and (b) samples B and C at room temperature. Open circles are data points, solid lines are fits to the data.

FIG. 3. Hanle curves measured in a 3-terminal geometry for a 100 μm x 150 μm Fe/MgO contact. (a) Room temperature data as a function of bias for injection (left axis) and extraction (right axis) from sample A, and (b) as a function of temperature at an injection bias of –0.5 V from sample A. Room temperature injection for several biases into (c) sample B and (d) sample C. Open circles are data points, and solid lines through the data are Lorentzian fits described in the text.

FIG. 4. Spin lifetime, $\tau_s$ (circles-left axis) and spin diffusion length, $\lambda_s$ (triangles -right axis) vs. carrier concentration for the Fe/MgO/Ge system.

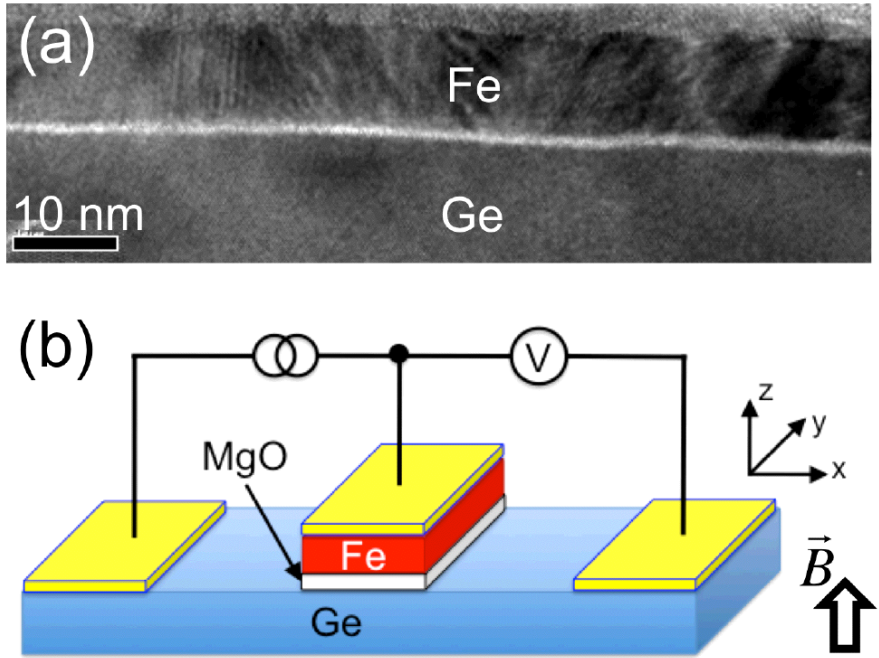

Figure 1 – Hanbicki, et al.

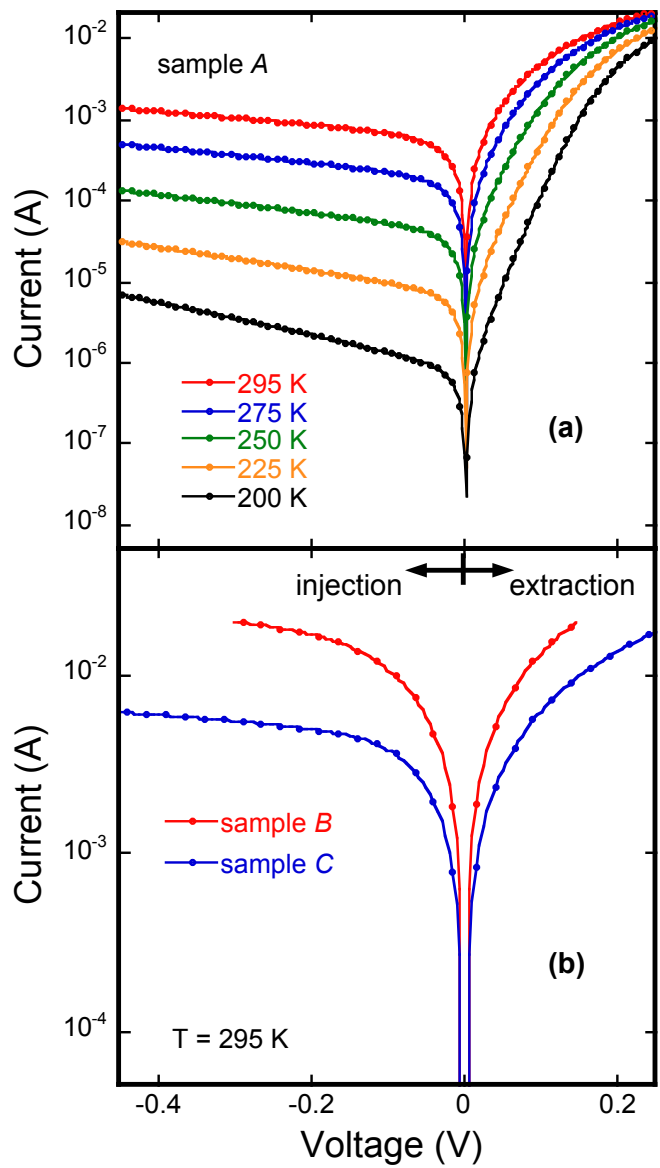

Figure 2 – Hanbicki, et al.

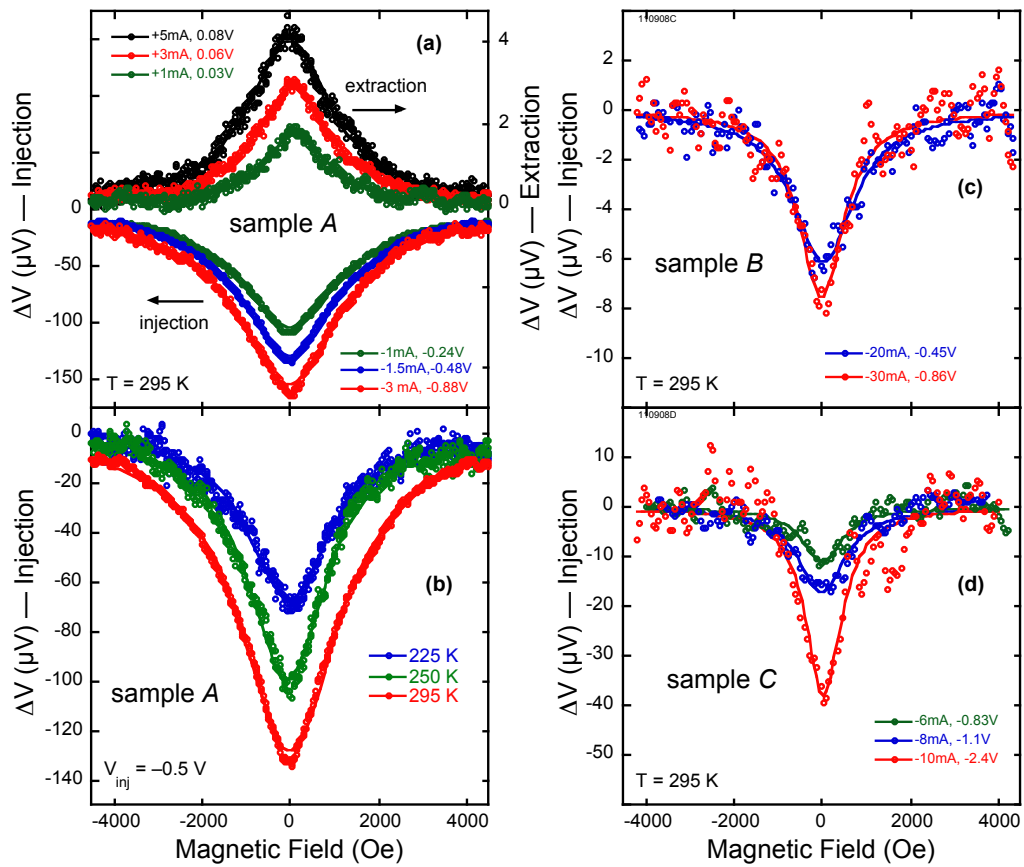

Figure 3 – Hanbicki, et al.

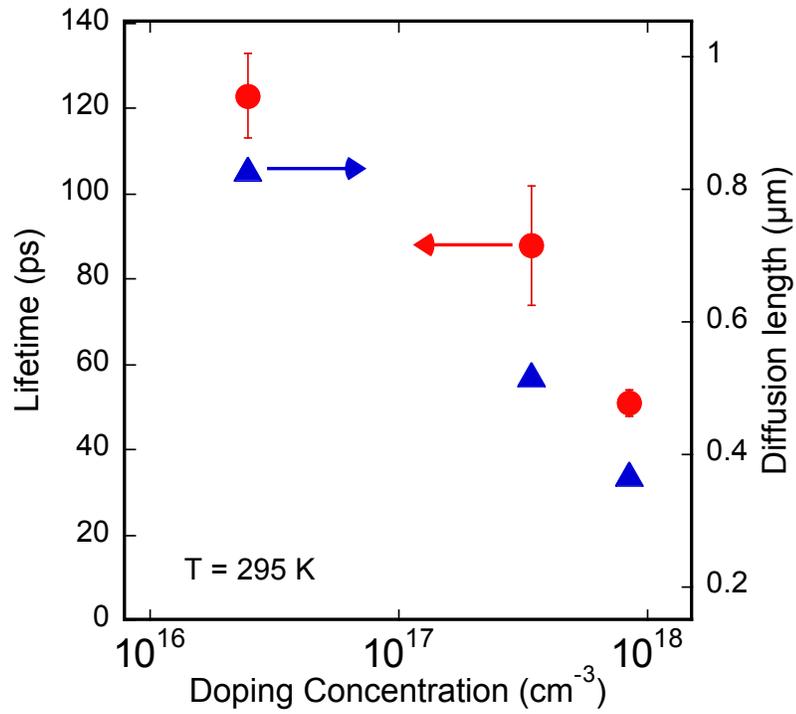

Figure 4 – Hanbicki, et al.